\documentclass[traditabstract]{aa} 
\usepackage[T1]{fontenc}
\usepackage[utf8,latin1]{inputenc}
\usepackage{epsfig}
\usepackage{graphicx}
\usepackage{txfonts}
\usepackage{amssymb}
 \usepackage{siunitx}
\usepackage{natbib}                               
\bibpunct{(}{)}{;}{a}{}{,}

\def\compps{{\sc compps}} 
 
\def\bbodyrad{{\sc bbodyrad}}

\def\Integ{{\em INTEGRAL}} 
\def\rxte{{\em RXTE}} 
\def\swift{{\em Swift}} 
\def\chandra{{\em Chandra}} 
 
\def\igr{IGR~J00291+5934} 
\def\be{\begin{equation}} 
\def\ee{\end{equation}} 
 
\begin{document} 

\title{The 2015 outburst of the accretion-powered pulsar \igr: \Integ\ and \swift\ observations} 
 
\author{V. De Falco\inst{1,2}
\and L. Kuiper\inst{3} 
\and E. Bozzo\inst{4}
\and D. K. Galloway\inst{5,6}
\and J. Poutanen\inst{7} 
\and C. Ferrigno\inst{4}
\and L. Stella\inst{8}           
\and M. Falanga\inst{2,1}
} 
 
\offprints{V. De Falco} 
 
\titlerunning{The 2015 outburst of the accretion-powered pulsar \igr}  
\authorrunning{De Falco et al.}
  
\institute{Institut f\"ur theoretische Physik, Universit\"at Basel, Klingelbergstrasse 82, 4056 Basel, Switzerland\\
\email{vittorio-df@issibern.ch}   
\and International Space Science Institute (ISSI), Hallerstrasse 6, 3012 Bern, Switzerland
\and SRON--Netherlands Institute for Space Research, Sorbonnelaan 2, 3584 CA, Utrecht, The Netherlands 
\and ISDC, Data centre for astrophysics, University of Geneva, Chemin d'\'Ecogia 16, 1290 Versoix, Switzerland
\and School of Physics and Astronomy, Monash University, VIC 3800, Australia
\and Monash Centre for Astrophysics, Monash University, VIC 3800, Australia
\and Tuorla Observatory, Department of Physics and Astronomy, University of Turku, V\"ais\"al\"antie 20, FI-21500 Piikki\"o, Finland
\and INAF -- Osservatorio Astronomico di Roma, via Frascati 33, Monteporzio Catone, Rome, 00078, Italy  
} 
 
\date{} 
 
\abstract{The pulsar \igr\ is the fastest-known accretion-powered X-ray pulsar, discovered during a transient outburst in 2004. In this paper, we report on \Integ\ and \swift\ observations during the 2015 outburst, which lasts for $\sim25$~d. The source has not been observed in outburst since 2008, suggesting that the long-term accretion rate has decreased by a factor of two since discovery. The averaged broad-band (0.1 -- 250~keV) persistent spectrum in 2015 is well described by a thermal Comptonization model with a column density of $N_{\rm H} \approx4\times10^{21}$ cm$^{-2}$, a plasma temperature of $kT_{\rm e} \approx50$ keV, and a Thomson optical depth of $\tau_{\rm T}\approx1$. Pulsations at the known spin period of the source are detected in the \Integ\ data up to the $\sim150$ keV energy band.  We also report on the discovery of the first thermonuclear burst observed from \igr, which lasts around 7 min and occurs at a persistent emission level corresponding to roughly $1.6\%$ of the Eddington accretion rate. The properties of the burst suggest it is powered primarily by helium ignited at a depth of $y_{\rm ign}\approx1.5\times10^9$ g cm$^{-2}$ following the exhaustion by steady burning of the accreted hydrogen. The Swift/BAT data from the first $\sim20$ s of the burst provide indications of a photospheric radius expansion phase. Assuming this is the case, we infer a source distance of $d = 4.2 \pm 0.5$ kpc.}

\keywords{pulsars: individual IGR~J00291+5934 -- stars: neutron -- X-ray: binaries -- X-ray: bursts}

\maketitle
 
\section{Introduction} 
\label{sec:intro}
Accreting millisecond X-ray pulsars (AMXPs) are old (in the order of Gyr) neutron stars (NSs) endowed with a relatively low magnetic field ($B\approx 10^{8-9}$ G) and spin frequencies typically between 180 -- 600 Hz. They are hosted in low mass X-ray binaries (LMXBs) characterised by an orbital period ranging from 40 min to 5 hr. All AMXPs are X-ray transients, spending most of their time in a quiescent state (X-ray luminosities of the order of $10^{31-32}$ erg s$^{-1}$) and sporadically undergoing outbursts that can last for a few weeks and reach X-ray luminosities of $10^{38}$~erg~s$^{-1}$ \citep[see, e.g.,][for reviews]{wijnands06,poutanen06,patruno12}. 

The AMXP \igr\ was discovered during an outburst in 2004 and it is the fastest spinning AMXP discovered so far. Its spin period is 1.67 ms \citep{markwardt04b,falanga05}. \igr\ is also the first AMXP showing a clear spin-up behaviour during outburst \citep{falanga05}. Together with other AMXPs, such as SAX~J1808.4--3658 \citep[the first discovered in 1998;][]{Wijnands98}, PSR~J1023+0038 \citep{shahbaz15}, and IGR~J18245--2452 \citep[][]{papitto13b}, it confirms the evolutionary link between the accretion powered LMXBs and the rotation-powered millisecond pulsars \citep[so-called recycling scenario;][]{Alpar82}. \rxte/ASM data suggest that possible outbursts from the source could have occurred already in 1998 and 2001 \citep{remillard04}, tentatively indicating a recurrence time of $\sim3$ yr. This possibility can not be confirmed further, as the following outbursts from \igr\ are recorded only in 2008 and 2015 \citep{lewis10,hartman11,lipunov15,sanna15}. The quiescent emission from the source is studied by \citet{jonker05}, who record with \chandra\ a luminosity as low as (5 -- 10)$\times10^{31}$~erg~s$^{-1}$ (0.5 -- 10 keV). This value is comparable with that observed from other AMXPs in quiescence. \citet{patruno10} also reports the possible spin-down between the outbursts that would be expected according to the standard magneto-dipole radiation scenario \citep[see, e.g.,][for a review]{patruno12}. 

The donor star hosted in \igr\ is identified to be a hot brown dwarf with a mass ranging between 0.039 -- 0.16 M$_\odot$ \citep{galloway05}. This is in agreement with stellar evolutionary expectations, which predict that the observed AMXPs with an orbital period in the 1.4 -- 11 hr range (that in our case is 2.5 hr) should host hydrogen-rich white or brown dwarfs \citep[see ][and references therein]{bildsten01}. Therefore, all these AMXPs also exhibit helium type-I X-ray bursts after they burned prior to ignition of the accreted hydrogen fuel \citep[see e.g.,][]{galloway06,watts06,falanga07,galloway07, falanga11,ferrigno11}. 

In this paper, we report on \Integ\ and \swift\ observational campaign carried out to monitor the \igr\ July 2015 outburst. We study the properties of the broad-band spectral energy distribution of the source, as well as its timing properties. We also present a detailed analysis of the first thermonuclear X-ray burst observed from \igr\ \citep[see also][]{kuin15,bozzo15}.   
 
\section{Observations and data} 
The 2015 outburst is covered by \Integ\ (22 -- 250 keV) and \swift\ (5 -- 22 keV) data (see Sect.~\ref{sec:integral} and ~\ref{sec:swift}). To have the full coverage of the outburst trend we have used all the daily-averaged available \swift/BAT data (15 -- 50 keV).

\subsection{INTEGRAL} 
\label{sec:integral}  

We analysed the \Integ\ \citep{w03} target of opportunity (ToO) observation performed on \igr\ during its outburst in July 2015. The observation started on 2015 July 27 at 17:45:57.2 UTC and ended on July 29 at 21:07:17.2 UTC, summing up to a total exposure time of 170 ks (see Table \ref{tab:swift}). 

The reduction of the \Integ\ data was performed using the standard {\sc offline science analysis (OSA)} version 10.2 distributed by the ISDC \citep{c03}. \Integ\ data were divided into science windows (ScW), that is, different pointings each lasting $\sim 2-3$\,ks. The algorithms used for the spatial and spectral analysis are described in \citet{gold03}. The observation, that is in a hexagonal dithering mode aimed at \igr, consisted of 50 ScWs with a source position offset $\lesssim 2\fdg5$ from the centre of the field of view.  We analysed data from the IBIS/ISGRI coded mask telescope \citep{u03,lebr03}, covering the 20 -- 300~keV energy band, and from the two JEM-X monitors \citep{lund03}, covering the 3 -- 20~keV energy range.

The pulsar \igr\ was detected in the IBIS/ISGRI mosaic at a significance level of $60\sigma$ (20 -- 100 keV energy range). We note  that the nearby source V709~Cas (detection significance of $9\sigma$ in the 20 -- 100 keV energy band) can be clearly distinguished thanks to the spatial resolution of the instrument and thus is not contaminating the X-ray emission recorded from \igr\ \citep[see also][]{falanga05}. \igr\ is also clearly detected in the JEM-X mosaics with a detection significance of $42\sigma$ in the 3 -- 20~keV energy range. The best determined position of \igr\ is at $\alpha_{\rm J2000} = 00^{\rm h} 29^{\rm m} 01\fs97$ and $\delta_{\rm J2000} = 59{\degr}34\arcmin18\farcs9$, with an associated uncertainty of $3\farcs5$ at the 90\% confidential level \citep[c.l.; 20 -- 100 keV;][]{gros03}. The offset with respect to the position of the associated optical counterpart is $0\farcs2$ \citep{torres08}. We first extracted a number of energy resolved light curves for IBIS/ISGRI and the two JEM-X, at a time scale of one ScW. The analysis of these light curves does not provide evidence for significant spectral variation during the outburst, and thus we extracted a single ISGRI and JEM-X spectrum averaged over the entire exposure time available. These spectra are described in the next section and fit together with the \swift\ data.  

\subsection{Swift} 
\label{sec:swift}  
\swift/XRT \citep{burrows05} started to monitor the source $\sim135$ s after a BAT trigger that occurred on 2015 July 25 at 02:12 UTC, which corresponds to the onset of the first type-I X-ray burst observed from this source \citep{kuin15}. The XRT monitoring campaign covers the source outburst from 57227~MJD to 57250~MJD, comprising a total of 13 pointings and an effective exposure time of 21~ks. One XRT observation is carried out on 2015 July 29 (ID~00031258006) simultaneously with the \Integ\ ToO. The effective exposure time of this XRT pointing is 1.4 ks (see Table \ref{tab:swift}). 

We processed the \swift/XRT data by using standard procedures \citep{burrows05} and the calibration files version 20160113. The \swift/XRT data awere taken both in window-timing (WT) and photon-counting (PC) modes (processed with the {\sc xrtpipeline} v.0.13.2). Filtering and screening criteria were applied by using FTOOLS contained in the {\sc heasoft} software package (v6.19)\footnote{http://heasarc.gsfc.nasa.gov/docs/software.html.}. We extract source and background light curves and spectra by selecting event grades of 0 -- 2 and 0 -- 12 for the WT and PC mode, respectively. We used the latest spectral redistribution matrices in the HEASARC calibration database. Ancillary response files, accounting for different extraction regions, vignetting and PSF corrections, were generated using the {\sc xrtmarkf} task. When required, we corrected PC data for pile-up, and used the {\sc xrtlccorr} task to account of this correction in the background-subtracted light curve.

We extracted the BAT light curve with {\sc batgrbproduct} tool and standard techniques and we also used the public daily-averaged \swift/BAT data in the 15 -- 50 keV energy band retrieved from the Hard X-ray Transient Monitor webpage\footnote{http://swift.gsfc.nasa.gov/results/transients/} \citep{krimm13}. 

\begin{table}[h] 
\caption{Log of all \swift\ and \Integ\ observations used in this work.}
\centering
\begin{tabular}{ccccc} 
\hline 
\hline 
Sequence & Obs/mode & Start time (UT) & Exposure\\
& & (UTC)  & (ks)\\
\hline 
00031258005 & \swift/WT & 2015-07-24 13:23:42 & 1\\ 
00650221000 & \swift/WT & 2015-07-25 01:56:19 & 0.3\\
00650221001 & \swift/WT & 2015-07-25 03:21:27 & 2\\
00031258006 & \swift/WT & 2015-07-29 00:18:11 & 1.4\\
00031258007 & \swift/WT & 2015-07-31 12:46:58 & 1.9\\
00031258008 & \swift/PC & 2015-08-02 01:27:58 & 2.5\\
00031258009 & \swift/PC & 2015-08-04 02:48:58 & 2\\
00031258010 & \swift/PC & 2015-08-06 18:45:58 & 2\\
00031258012 & \swift/PC & 2015-08-09 08:57:57 & 1.9\\
00031258014 & \swift/PC & 2015-08-14 16:57:57 & 1.8\\
00031258015 & \swift/PC & 2015-08-17 08:38:12 & 0.9\\
00031258016 & \swift/PC & 2015-08-19 11:31:58 & 1.1\\
00031258017 & \swift/PC & 2015-08-23 10:00:58 & 1.7\\
\hline  
\hline  
15690004001 & \Integ & 2015-07-27 17:45:57 & 170\\
\hline  
\end{tabular} 
\label{tab:swift}
\end{table}

\section{Outburst properties} 
\subsection{The light curve} 
\label{sec:lc} 
In Fig. 1 we show the light curve of \igr\ as obtained from all data used in this paper and showing the entire source outburst. The count-rates measured from all instruments were converted to bolometric flux (0.1 -- 250 keV) using the spectral analysis results from Sect.~\ref{sec:specIGRJ}. 

The source displays a clear increase in flux for $\sim3$~d after the onset of the event, reaching a peak value that is about three orders of magnitude larger than the persistent quiescent flux recorded before 57225~MJD. The source then decays back into its quiescent state around 57250~MJD, suggesting that the outburst lasts $\sim25$~d in total. 
\begin{figure}[h] 
\centering 
\centerline{\epsfig{file=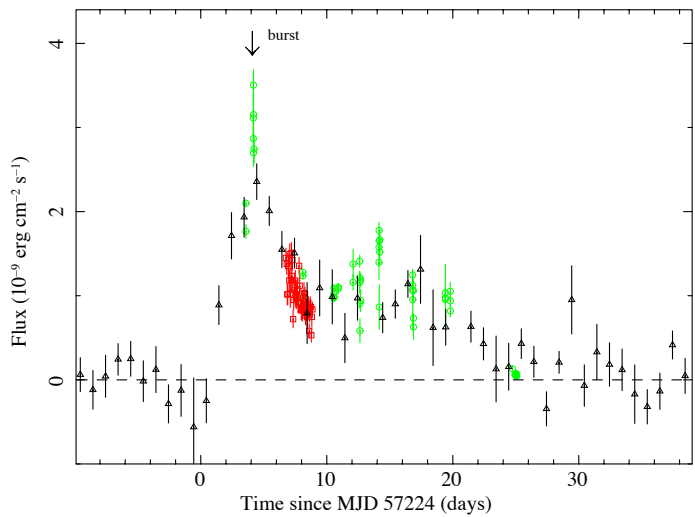, trim=3.5cm 1cm 5cm 2.5cm, scale=0.38}} 
\caption{Light curve of the 2015 outburst observed from \igr.\ We show data from \swift/XRT 
(green circles, bin time of 500~s; July 24 -- 15 August), \swift/BAT (black triangles, each measurement is averaged over 1~d; July 11 -- 28 August), and \Integ/ISGRI (red squares, the integration time is one science window of $\sim2$ -- 3 ks; July 27 -- 29). The arrow indicates the time of the only detected type-I X-ray burst. The dashed line shows the quiescent flux level. The source outburst reaches the highest peak flux that has ever been detected in the previous outbursts.}
\label{fig:fig1} 
\end{figure}

The profile of the outburst over time, shown by the source in 2015, is strongly reminiscent of that recorded during the event in 2004, as in both cases the return to quiescence occurs on a time scale of 15 -- 25~d and also the peak flux is roughly comparable \citep[we note that the onset of the outburst in 2004 can not be observed; see][and references therein]{falanga05}. The profile of the source outburst corresponding to the 2008 event is much more peculiar, featuring a double outburst light curve separated by 30~d. On that occasion, the source displayed a clear rise and fast decay in flux during the first 9 d of the event and then remains below the instrument detection level for about a month. A second brightening phase follows, lasting roughly $\sim 15$ d \citep{hartman11}. The origin of this singular outburst profile is still a matter of debate \citep[see also the discussion in][]{bozzo15}. The analysis of the long term BAT light curve shows that no rebrightening episode takes place after the main outburst in 2015. 

A novel feature of the 2015 outburst of \igr\ is the first detection of a thermonuclear burst, occurring 2.8~d after the event onset, close to the time of the peak flux achieved by the source \citep{kuin15,bozzo15}. We perform a time-resolved spectral analysis of the source X-ray emission during the type-I burst in Sect.~\ref{sec:burst}. No additional thermonuclear bursts are found either in the continuous 170~ks \Integ\ monitoring observation or in the available 13 \swift/XRT pointings lasting 21 ks.

\subsection{The recurrence time}
\label{sec:dg} 
Prior to the 2015 outburst, \igr\ has exhibited outbursts roughly every three years, with a gradually increasing interval (Table \ref{tab:outbursts}). The regularity of the outbursts allows the time of the 2008 outburst to be predicted to within 1\% of the actual interval, 3.7~yr \citep{galloway08}. The same quadratic fit would predict that the next outburst should occur after 4.1~yr, around 56200~MJD (2012 October); instead, it occurs some 2.8~yr later, after an interval of 6.9~yr. 

It is tempting to speculate that the pair of outbursts observed in 2008, separated by one month, may exhaust the disk to an unusual extent, delaying the following outburst. In fact, the total fluence of both 2008 outbursts together is slightly below that of the 2004 outburst. The 2015 outburst is somewhat more energetic again; the integrated flux from the {\it Swift}/BAT and {\it INTEGRAL}/ISGRI measurements indicates a total fluence of $(2.01\pm0.11)\times10^{-3}\ {\rm erg\,cm^{-2}}$. Furthermore, it is usually assumed that such transient outbursts exhaust all the accumulated material in the disk (although admittedly the secondary 2008 outburst is evidence that this is not always the case). We can also rule out any missed outbursts between 2008 and 2015, based on {\it Swift}\/ BAT monitoring. Due to the high declination of the \swift\ satellite, \igr\ is constantly monitored by BAT on a daily timescale, and the typical upper limit outside the outburst is $5\times10^{-10}\ {\rm erg\,cm^{-2}\,s^{-1}}$ (priv. comm., H. Krimm, 2016).

\begin{figure}[h] 
\centering 
\centerline{\epsfig{file=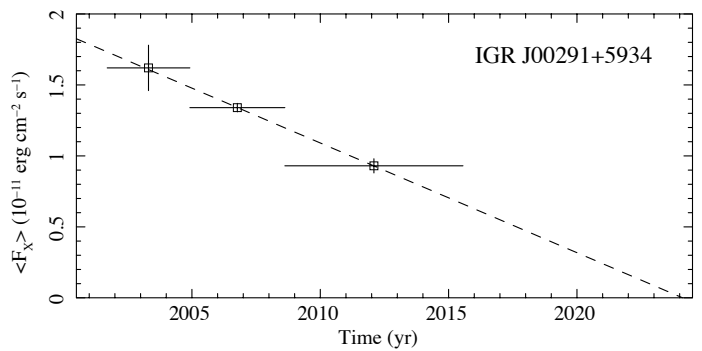, trim=2.8cm 9cm 5cm 0.2cm, scale=0.38}} 
\caption{Estimated time-averaged X-ray flux for \igr\ for the outbursts observed between 2001 -- 2015, based on the outburst fluences listed in Table \ref{tab:outbursts}. The 2008 August and September outbursts are taken together to estimate the average rate since 2004. The average flux can be taken as a measure of the accretion rate into the disk between outbursts; the dashed line shows a linear fit, which (extrapolated) decreased to zero by 2024. We predict that the source may not exhibit another outburst as strong as that in July 2015 in the next decade.}
\label{fig:fig6} 
\end{figure}

Instead, we find the long-term outburst history offers strong evidence that the steady mass transfer rate is decreasing, by about 50\% between 2001 and 2015. We calculate in Table \ref{tab:outbursts} the time-averaged X-ray flux as the outburst fluence divided by the outburst interval. This value is $1.6\times10^{-11}\ {\rm erg\,cm^{-2}\,s^{-1}}$ between 2001 -- 2004, but only $9.3\times10^{-12}\ {\rm erg\,cm^{-2}\,s^{-1}}$ between 2008 -- 2015 (see Fig.~\ref{fig:fig6}). 
For a distance of 4~kpc, this corresponds to a range of accretion rates of (0.2 -- 0.01)\%~$\dot{M}_{\rm Edd}$. Such a decrease in the long-term accretion rate is also inferred from the outbursts in SAX~J1808.4$-$3658, on a similar timescale \citep{galloway08}. Following the same line of thought of \citet{galloway08}, we also perform a linear fit to the average fluxes over the last three outburst intervals. It may indicate that the time-averaged X-ray flux has dropped steadily with the rate of $\sim0.08\times10^{-11}\ {\rm erg\,cm^{-2}\,s^{-1}\,yr^{-1}}$. While this fit is purely phenomenological, if the trend continues, the wait time for future outbursts (of similar fluence to those in 2004 and 2015) will occur substantially longer than seven years.
\begin{table*}[t] 
\caption{Transient outbursts from \igr.}
\centering
\scalebox{1.2}{
\begin{tabular}{ccccccccc} 
\hline 
\hline 
Outburst start & Duration & Interval\tablefootmark{a} & $F_{\rm peak,bol}$\tablefootmark{b} & Fluence\tablefootmark{c} & $\left<F_X\right>$\tablefootmark{d}& Refs.\tablefootmark{e}\\
calendar date & d & yr & $10^{-9}\ {\rm erg\,cm^{-2}\,s^{-1}}$ & $10^{-3}\ {\rm erg\,cm^{-2}}$ & $10^{-11}\ {\rm erg\,cm^{-2}\,s^{-1}}$ \\
\hline 
26 Nov 1998 & -- & $>2.9$ & -- & -- & ($<1.8$) & -- \\
11 Sep 2001 & -- & 2.8 & -- & -- & (1.8) & --\\ 
2 Dec 2004  & 14 & 3.2 & $2.9\pm0.2$ & $1.6\pm0.2$ & $1.6\pm0.2$ & [1] \\ 
13 Aug 2008  & 9  & 3.7 & $1.5\pm0.2$ & $0.76\pm0.03$ & $0.65\pm0.03$ & [2]\\
18 Sep 2008  & 15 & 0.1 & $1.1\pm0.1$ & $0.79\pm0.03$ & $25\pm1$ & [2]\\
22 Jul 2015  & 25 & 6.9 & $3.5\pm0.2$ & $2\pm0.1$ & $0.9\pm0.1$ & [3]\\
\hline  
\end{tabular} } 
\tablefoot{
\tablefoottext{a}{The epoch for the outburst prior to the first known is assumed to be earlier than the first \rxte/ASM measurements (typically 1996 January 6 or 50088~MJD).}
\tablefoottext{b}{The peak bolometric flux is in the 0.1 -- 250 keV energy range.}
\tablefoottext{c}{Bolometric fluence.}
\tablefoottext{d}{Estimated time-averaged bolometric flux. Values in parentheses are estimated from the \rxte/ASM intensity, since \rxte/PCA measurements are not available at the outburst peak, and are approximate.}
\tablefoottext{e}{References: [1] \citet{galloway05}; [2] \citet{hartman11}; [3] this paper.}
}
\label{tab:outbursts}
\end{table*} 

\section{Spectral analysis} 
\label{sec:specIGRJ}
The spectral analysis is carried out using {\sc xspec} version 12.6 \citep{arnaud96}. All uncertainties in the spectral parameters are given at a $1\sigma$ confidence level for a single parameter. 

We first fitted all the different XRT spectra extracted from the available 13 pointings, excluding the type-I burst. 
This analysis revealed that in all cases the source X-ray emission in the soft energy band can be reasonably well 
described by using a simple absorbed power-law model ($\chi^{2}_{\rm red}{\rm /d.o.f.}= 0.98/97$). In all cases, we measured an absorption column density of $N_{\rm H}=(1.30\pm0.15)\times10^{22}$~cm$^{-2}$, and a photon index of $\Gamma=1.96\pm0.05$, with no evidence of significant spectral variability. We thus performed a more detailed broad-band spectral fit by combining the XRT observations carried out on 2015 July 29 with the \Integ\ data. In order to limit the uncertainties in the calibrations of the different instruments, the fit to the broad-band spectrum is limited to 1.1 -- 7.5 keV for the XRT data in WT mode, 5 -- 22 keV for the two JEM-X and 22 -- 250 keV for ISGRI. A constant factor was included in the fit in order to take into account the inter-calibrations between the different instruments and the possible intrinsic variability of the source (in all cases we assume ISGRI as the reference instrument and fix the corresponding constant to unity). 

\begin{figure}[h] 
\centering 
\centerline{\epsfig{file=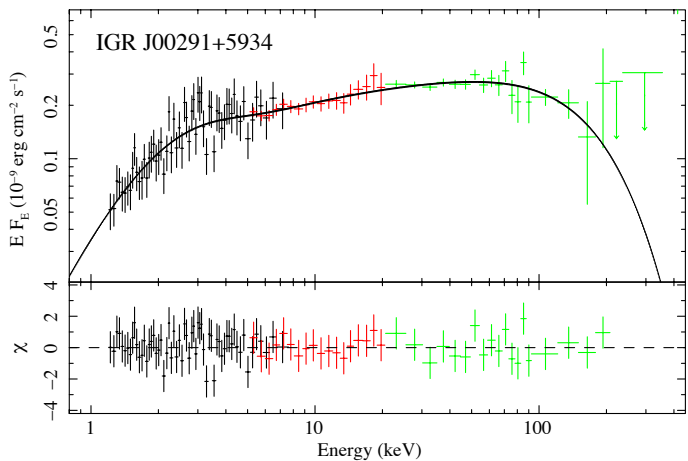, trim=3cm 1.5cm 5cm 4cm, scale=0.38}} 
\caption{Unfolded measured broad-band spectrum of \igr\ as observed by JEM-X (red points), ISGRI (green points) and \swift/XRT (black points). The best fit is obtained with the \compps\ model with a plasma temperature $kT\approx50$ keV, which is represented in the figure with a solid black line. The residuals from the best fit are shown in the bottom panel.}
\label{fig:fig2} 
\end{figure}

We obtained an acceptable fit to the broad-band spectrum of \igr\  using an absorbed cut-off power-law model ($\chi^{2}_{\rm red}{\rm /d.o.f.}= 0.73/95$). We measured a column density $N_{\rm H}=(1.01\pm0.17)\times10^{22}$~cm$^{-2}$, a photon index of $\Gamma=1.7\pm0.1$ and a cutoff energy of $E_{\rm cut}=147^{+77}_{-39}$ keV. In order to achieve a more physical description of the source X-ray emission, we replaced the cutoff power-law model with the thermal Comptonization model \compps,\ under the assumption of a slab geometry \citep{ps96}. This model has been successfully applied in the past to a number of other AMXPs \citep[see e.g.,][]{gp05,mfb05,falanga05,mfc07,ip09,bozzo10,falanga11}. The new set of model parameters are the absorption column density $N_{\rm H}$, the plasma temperature of the accretion column $kT_{\rm e}$, the blackbody temperature $kT_{\rm bb}$ of the soft-seed photons assumed to be injected from the bottom of the slab, the Thomson optical depth $\tau_{\rm T}$ across the slab, and the inclination angle $\theta$ between the slab normal and the line of sight. This model provides a fully acceptable result ($\chi^2_{\rm red}/{\rm d.o.f.}=0.72/102$) and we report all values of the best fit parameters in Table \ref{tab:table1}.  
\begin{table}[h] 
\caption{\label{table:spec} Best parameters determined from the fit to the broad-band spectrum of \igr\ with the 
\compps\ model.}
\centering
\scalebox{1}{
\begin{tabular}{ll} 
\hline 
& \compps\ \\
\hline 
$N_{\rm H}\ (10^{21}\ {\rm cm}^{-2})$ & $4_{-3}^{+7}$\\ 
$kT_{\rm e}$ (keV) & $49 \pm 12$\\ 
$kT_{\rm bb}$ (keV) & $0.6^{+0.2}_{-0.1}$\\ 
$\tau_{\rm T}$ & $1.4^{+0.5}_{-0.3}$\\ 
$\cos \theta $ & $0.6\pm0.2$\\
$A_{\rm seed}$\tablefootmark{a} (km$^2$) & $48\pm15$ \\ 
$\chi^{2}_{\rm red}/{\rm d.o.f.}$ & 0.72/102 \\
$L_{\rm bol}$\tablefootmark{a} ($10^{37}$ erg s$^{-1}$) & 0.21$\pm$0.02\\
\hline  
\end{tabular}}  
\tablefoot{ \tablefoottext{a}{Assuming a source distance of 4 kpc (see Sect.~\ref{sec:burst}).}}
\label{tab:table1} 
\end{table} 
The column density $N_{\rm H}=(4^{+7}_{-3})\times10^{21}$~cm$^{-2}$ turns out to be fully in agreement (to within the uncertainties) with that estimated previously $N_{\rm H}\approx2.8\times10^{21}\  {\rm cm}^{-2}$ from the {\it Chandra} and \swift\ observations \citep{nowak04,torres08,bozzo15}. The \compps\ model also allows us to determine the apparent area of the hot spot region on the NS surface, $A_{\rm seed}\approx 48\ (d/4 \mbox{kpc})$ km$^2$. At the estimated distance of \igr\ (see Sect.~\ref{sec:burst}), the radius of the hot spot is $\approx 4$ km. We note that the spectral parameters reported in Table \ref{tab:table1} are comparable to those measured by \citet{falanga05} during the source outburst in 2004. Figure~\ref{fig:fig2} shows the unfolded measured broad-band spectrum together with the residuals from the best fit model. In this fit, the normalization constants of the JEM-X and XRT data are 1.02 and 1.8, respectively compared to the ISGRI data fixed at unity. The large variability of the multiplicative normalization factor between XRT and ISGRI data may be due to the different exposure time. In addition, the XRT pointing is coincident with a small flare.

\section{Timing analysis}
\label{sec:tmchar}
We study the ephemeris and the pulse profile of \igr\ in the 20 --- 150~keV energy range taking advantage of the relatively good statistics of the ISGRI data. For an AMXP timing analysis we have insufficient time-resolution both in \swift/XRT PC and WT modes. The timing accuracy of the ISGRI time stamps recorded on-board is about $61\mu$s. We convert all ISGRI on-board arrival times from the terrestrial time (TT) reference system to the solar system barycenter (TDB). This barycentering procedure requires an accurate knowledge of the instantaneous spacecraft position and velocity, the best known (optical) pulsar position \citep{torres08}, and the solar system ephemeris (DE200) information. We excluded from this analysis events recorded during time periods where the ISGRI count rate behaves erratically (e.g., near perigeum ingress and egress, or during high solar activity). We further selected only events with rise times between channels 7 -- 90 \citep{lebrun03} from non-noisy pixels having a pixel illumination factor larger than 25\%.

The spin period of \igr\ is known to evolve in time \citep[]{patruno10,hartman11,papitto11} and thus the precise rotational frequency of the pulsar, as well as the correspondingly updated binary orbital parameters, can not be accurately predicted a priori for the outburst of 2015. We thus first performed a search for coherently modulated signals in the ISGRI data around the most probable pulsar rotational frequency, assuming binary parameters reasonably similar to previously reported values. In order to maximise the signal to noise ratio of the data we additionally screened out events that are outside the energy range 20 -- 60 keV \citep[see discussion in][]{falanga05}. The systematic search was carried out by using a code based on a {\sc simplex} optimisation scheme, that finds the global minimum of the $Z_1^2$-test statistics  \citep{buccheri1983} as a function of the spin frequency and the time of the ascending node, $T_{\rm asc}$. The arrival time of each event is first corrected for the binary motion (thus involving the time of ascending node) and subsequently converted into a pulse phase using the assumed spin frequency (the other free parameter). The best parameters determined with this technique are $\nu_{\rm spin}=598.8921299(8)$~Hz at epoch 57231.0 MJD (TDB) and $T_{\rm asc}=57231.847035(25)$~MJD (TDB) \citep{kuiper15} with a detection significance of pulsations of $5.2\sigma$ (uncertainties are given at 1$\sigma$ confidence level). 

The spin frequency measured by ISGRI in the outburst in 2015 is consistent, to within the relatively large uncertainties, with the value that can be predicted from the previous observations in 2008 and taking into account any of the spin down torques reported by \citet{patruno10}, \citet{hartman11}, and \citet{papitto11}. The updated value obtained for $T_{\rm asc}$ also allows us to refine our knowledge on the system orbital period by fitting together all available measurements of the times of the ascending node. With this technique we obtain  $P_{\rm orb}=8844.07672(2)$~s. We find no evidence for a change in the orbital period and the corresponding $2\sigma$ upper limit on the orbital period derivative is $\dot{P}_{\rm orb}=1.1\times 10^{-12}$ s\ s$^{-1}$.
\begin{table}[h] 
\caption{\label{table:burst} Ephemeris of \igr\ obtained from the \Integ\ observations.}
\centering
\begin{tabular}{ll} 
\hline 
Parameter & Value \\
\hline 
Epoch data start/end (MJD) & 57230.7 -- 57232.9\\
Frequency & 598.8921299(8) Hz\\
Solar system ephemeris & DE200\\
Epoch of the period (MJD;TDB) & 57231.0\\
Orbital period & 8844.07672(2) s\\
$a_x \sin i$ \tablefootmark{(*)} & $64.990(1)$ (lt-ms)\\
Time of ascending node (MJD;TDB) & 57231.847035(25)\\
\hline  
\end{tabular}  
\tablefoot{ \tablefoottext{*}{From \citet{papitto11}.}}
\label{tab:table2} 
\end{table}  
\begin{figure}
  \begin{center}
  \centerline{\epsfig{file=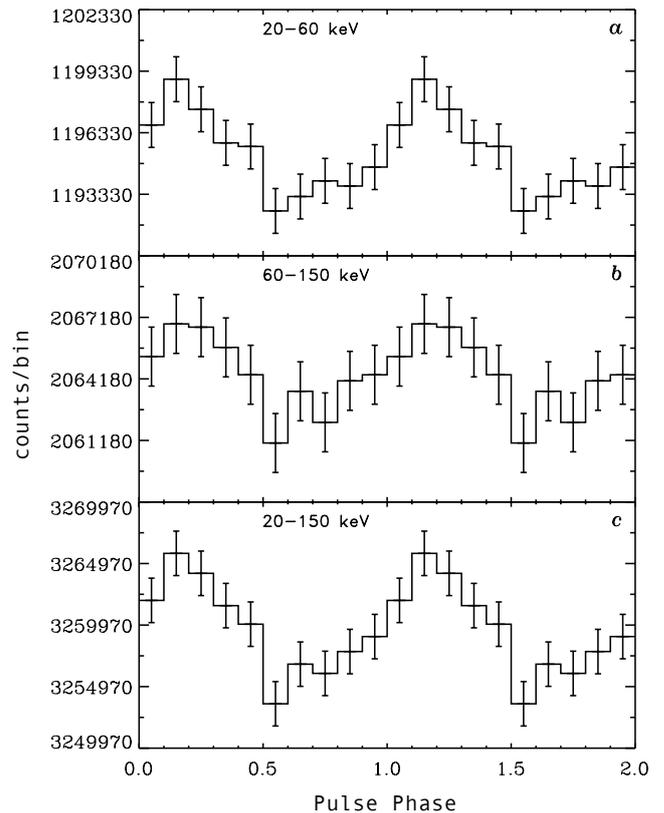, trim=3cm 5.5cm 5cm 5.5cm, scale=0.68}} 
     \caption{\label{pprofiles} Pulse profile of \igr\ in different energy bands during its 2015 outburst. 
     The pulsed emission is detected significantly by \Integ\ up to the $\sim150$ keV energy band.}
  \end{center}
   \label{fig:fig5} 
\end{figure}

We phase fold the barycentered ISGRI times upon the updated 2015 ephemeris, given in Table \ref{table:burst}. The resulting pulse-phase distributions are shown in Fig.~\ref{pprofiles} for three different energy bands. The $Z_1^2$-significances are $4.6\sigma, 3\sigma$ and $5.4\sigma$ for the 20 -- 60 keV, 60 -- 150 keV and 20 -- 150 keV bands, respectively. Thus, significant pulsed emission is detected for energies above 60 keV. We inspected the significance of the pulsed signal using the $Z_1^2$ statistics above 60 keV by progressively extending the energy range from 60 -- 90 keV to 60 -- 180 keV in steps of 30 keV. We find that the significance increased as function of the upper integration limit up to 150 keV and then flatten. This is consistent with the source emission being pulsed in the full energy range in which it is significantly detected.

\section{Properties of the type-I burst} 
\label{sec:burst} 
\begin{figure}[h] 
\centering 
\psfig{figure=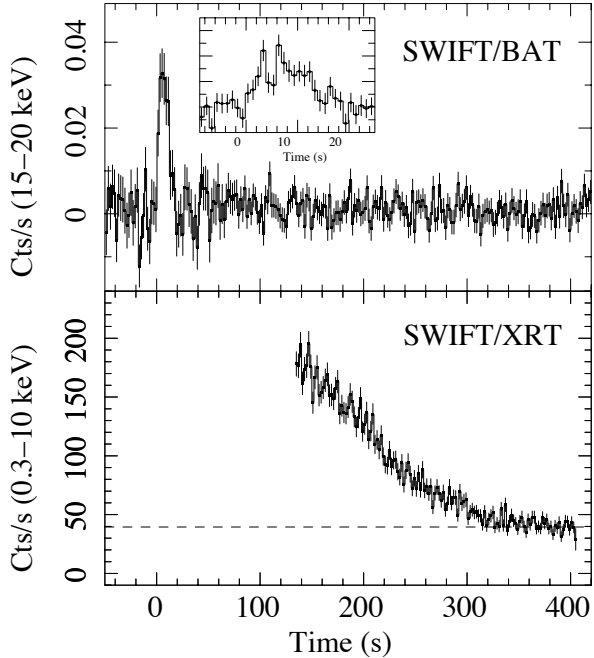, trim=1.5cm 1.8cm 3cm 4.35cm, scale=0.6}
\caption{Type-I burst detected during its 2015 outburst. The burst start time is 57228.0926~MJD. The upper panel shows the BAT light curve in the 15 -- 25~keV energy range with a time bin of 2~s (1~s for the inset). The XRT light curve in the bottom panel shows that the narrow field instrument on-board \swift\ starts to observe the source $\sim135$ s after the onset of the event. The dashed line in this panel indicates the emission level of the X-ray background as measured $\sim 400$ s after the burst. The burst is very energetic during the first $\sim20$ s with the tail of the blackbody detected above 25 keV.}
\label{fig:fig3} 
\end{figure}

In Fig.~\ref{fig:fig3}, we show the \swift/BAT 15 -- 20 keV (upper panel) and XRT/WT 0.3 -- 10 keV (lower panel) 
light curves of the only type-I burst observed so far from \igr.\ The burst start time of 57228.0926~MJD corresponds to the point at which the X-ray intensity of the source in BAT increases by 15\% with respect to the persistent level. The two plots in the figure suggest that the type-I burst is very energetic during the first $\sim 20$~s, with the tail of the black-body emission becoming detectable above 25 keV. The WT light curve starts with a delay of $\sim135$ s after the onset of the burst caught by BAT. The total duration of the burst, that is, the time to evolve away from and return to the persistent state, is $\sim7$ min. 

\begin{figure}[h] 
\centering 
\psfig{figure=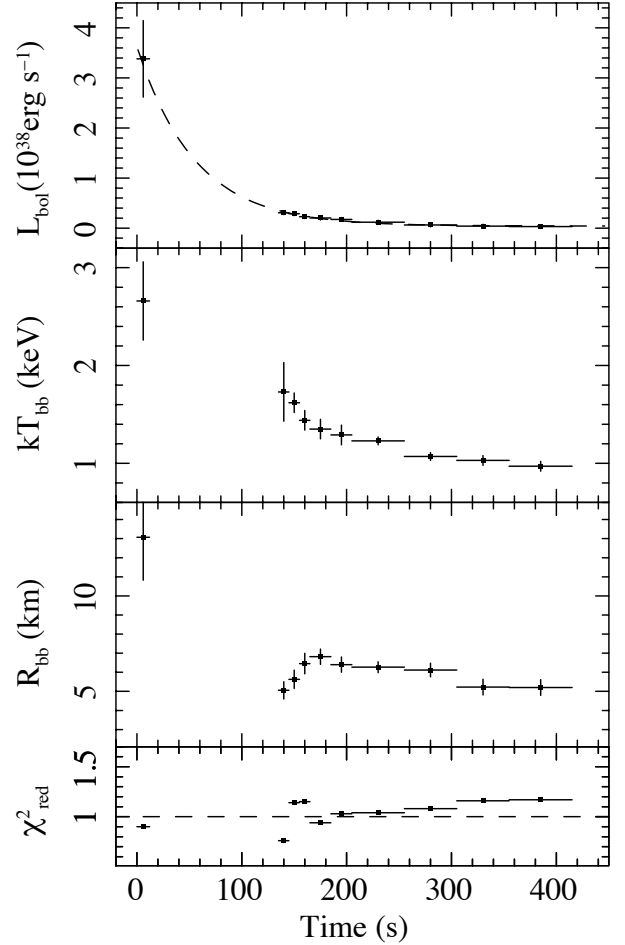, trim=2cm 2.6cm 5cm 2.2cm, scale=0.76}
\caption{Evolution of the spectral parameters during the type-I burst, as measured from BAT (first point) and XRT. 
The bolometric luminosity and the \bbodyrad\ radius are computed by using a distance of 4 kpc. The bottom panel shows the $\chi^2_{\rm red}$ values for all fits. The dashed line in the first upper panel represents the best determined exponential function with which we fit the burst profile. It shows an e-folding time of $\tau_{\rm fit}\approx55$ s.}
\label{fig:fig4} 
\end{figure}

To perform a time-resolved spectral analysis of the burst, we extract 9 XRT spectra during the event. All these spectra can be well fitted by using an absorbed black-body model (\bbodyrad\ in XSPEC) with the absorption column density fixed at the value determined from the broad-band analysis (i.e., $4\times10^{21}$ cm${}^{-2}$; see Table~\ref{tab:table1}). However, if we leave the $N_{\rm H}$ value free to vary, we find a value consistent with $\approx4\times10^{21}$ cm${}^{-2}$. We plot in Fig.~\ref{fig:fig4} the measured values of the apparent radius, $R_{\rm bb}$, the colour temperature, $kT_{\rm bb}$, and the bolometric luminosity, $L_{\rm bol}$, of the thermal emitting region. 

The luminosity at the peak of the burst, $F_{\rm peak}$, is obtained from BAT. The uncertainty derived from the spectral fit is increased by 20\% as the energy range of the instrument is largely outside the peak energy of the black-body \citep[see for more details Fig. 4 in][and \citet{bozzo15a}]{falanga09}. Taking into account the errors on the $N_{\rm H}$ value, this does not change the 20\% uncertainty on the burst peak flux. All the other luminosities are obtained from the XRT fluxes extrapolated in the 0.1 -- 35 keV energy range by generating ideal responses.
\begin{table}[h] 
\caption{\label{table:burst} Parameters of the type-I burst observed from \igr\ during its 2015 outburst.}
\centering
\begin{tabular}{ll} 
\hline 
\hline 
$F_{\rm peak}$\tablefootmark{a} ($10^{-8}$ erg cm$^{-2}$ s$^{-1}$) & $18\pm4$\\
$f_{\rm b}$\tablefootmark{b} ($10^{-6}$ erg cm$^{-2}$) & $10.7\pm2.6$\\
$F_{\rm pers,bol}$ ($10^{-9}$ erg cm$^{-2}$ s$^{-1}$) & $3.1\pm1.5$\\
$\tau_{\rm theo}\equiv f_{\rm b}/F_{\rm peak}$ (s) & $60\pm15$\\
$\gamma\equiv F_{\rm pers}/F_{\rm peak}$ & $(17.4\pm1.0)\times10^{-2}$\\
\hline  
\end{tabular}  
\tablefoot{ \tablefoottext{a}{Unabsorbed flux (0.1 -- 35 keV).} 
\tablefoottext{b}{Net burst fluence (0.1 -- 35 keV).}
}
\label{tab:table3} 
\end{table} 

We fitted the burst decay with an exponential function and derive an e-folding time of $\tau_{\rm fit}=55\pm11$ s (see Fig.~\ref{fig:fig4}). This is similar to the value obtained as $\tau_{\rm theo}\equiv f_{\rm b}/F_{\rm peak}=60\pm15$ s. In Table~\ref{tab:table3}, we report the measured burst parameters. The estimated burst fluence is $f_{\rm b}=E_{\rm burst}/(4\pi d^2)=1.1\times 10^{-5}$ erg cm${}^{-2}$, which corresponds to the total energy release of $E_{\rm burst}=2.1\times 10^{40}$ erg (at a source distance of $d = 4$ kpc, see below). 

The BAT burst light curve rebinned at 1s (see the small inset in Fig.~\ref{fig:fig3}) shows a double-peaked profile as evidence for a photospheric radius expansion \citep[PRE; see e.g.,][]{falanga07}. Assuming that the peak X-ray luminosity of the burst corresponds to the Eddington value $L_{\rm Edd}= 3.8 \times 10^{38}$ erg s${}^{-1}$ \citep[as empirically derived by][]{kuulkers03}, we can thus estimate the source distance. We obtain $d=4.2\pm0.5$ kpc in case of a pure helium burst or $d=2.7\pm0.5$ kpc for a thermonuclear burst ignited in material with a solar composition ($X_0=0.7$). For comparison, we note that the theoretical value of the source distance calculated by assuming a helium atmosphere and canonical NS parameters ($M_{\rm NS}=1.4M_\odot$ and $R_{\rm NS}=10$ km) would be $d=3.6\pm0.5$ kpc \citep{lewin93}. The estimated error on the distance is similar to those reported, for example, in \citet{kuulkers03}.

The amount of fuel liberated during the thermonuclear burning can be estimated as $E_{\rm burst}/\varepsilon_{\rm He}$, where $E_{\rm burst}=2.1\times 10^{40}$ erg is the observed total energy released during the burst and $\varepsilon_{\rm He} \approx 1.6\ \mbox{MeV/nucleon} \approx1.6 \times 10^{18}$ erg g$^{-1}$ is the total available nuclear energy for the transformation of pure helium into iron-peak elements. We thus obtain $E_{\rm burst} /\varepsilon_{\rm He}  \approx 1.3 \times 10^{22}$ g. This value is consistent with the estimated total mass $\Delta M$ accreted from the onset of the outburst to the start time of the thermonuclear burst ($\Delta t= 2.52641$ d). $\Delta M = \dot{M} \Delta t \approx 1.3 \times 10^{22}$ g, where $\dot{M}$ is derived from the X-ray flux at a distance of 4~kpc and the usual accreting equation $L_{\rm X}=GM_{\rm NS}\dot{M}/R_{\rm NS}$ \citep[see, e.g.,][]{frank02}. For a hydrogen type-I burst, the fuel burned during the thermonuclear burst would be $\approx 2.6 \times 10^{21}$~g, which is one order of magnitude lower than the measured total accreted matter ($\approx 1.2 \times 10^{22}$~g at $d=2.7$~kpc). The energetics of this burst can thus be explained by assuming that it is a helium type-I burst triggered by unstable helium burning, after any accreted hydrogen is exhausted by steady burning prior to the burst.
We thus assume $d\approx4$~kpc as the fiducial distance to \igr.\ 

At the distance of 4 kpc, the pre-burst persistent unabsorbed flux $F_{\rm pers,bol}\approx 3.1\times10^{-9}$ erg cm${}^{-2}$ s${}^{-1}$ translates into a bolometric luminosity of $L_{\rm pers,bol}\approx 6\times10^{36}$ erg s${}^{-1}$, or $\approx 1.6\%\ L_{\rm Edd}$. The local accretion rate per unit area is then given by $ \dot{m} = L_{\rm pers} (1+z) (4\pi R^2(GM/R))^{-1}$, that is, $\dot m \approx 3.3\times10^3$ g cm$^{-2}$ s$^{-1}$ (where the gravitational redshift is $1+ z = 1.31$). At this local accretion rate, helium type-I bursts are expected following completion of hydrogen burning, thus providing additional support to our conclusion above \citep[e.g., ][]{strohmayer2006}. 

The observed energy of the type-I burst allows us to estimate the ignition depth by using the equation $y_{\rm ign} = E_{\rm burst} (1+z)(4\pi R^2Q_{\rm nuc})^{-1}=1.5\times10^9$ g cm${}^{-2}$, where the nuclear energy generated for helium abundances (assuming a mean hydrogen mass fraction at ignition $\langle X\rangle=0$) is $Q_{\rm nuc}\approx 1.6+4\langle X\rangle\ {\rm MeV/nucleon}\approx1.6\ {\rm MeV/nucleon}$ \citep[][and references therein]{galloway04}, including losses owing to neutrino emission following \citet{fujimoto87}. The ignition depth is a particularly interesting parameter, as it regulates the recurrence time between different bursts through the equation $t_{\rm rec} =(y_{\rm ign} /\dot{m})(1 + z)$. At the mass accretion rate corresponding to the peak of the 2015 outburst shown by \igr\ ($1.6\%\ L_{\rm Edd}$ for pure helium), the expected recurrence time is $\sim 7$ d. Since the outburst from \igr\ lasts less than a month and the continuous coverage provided by \Integ\ is only 2.2 d, there is a high probability of missing another type-I burst. For a burst recurrence time of 7 d and for a total BAT exposure time of $\sim2$ ks, the probability of observing a burst is $0.3\%$.

\section{Summary and discussion} 
\label{sec:summary} 
We have studied the spectral and the timing behaviours of \igr\ during its outburst in 2015 by using the available \Integ\ and \swift\ observations. We have discussed the outburst recurrence time, that after the last three outburst intervals will drop to zero 8.4 yr after the July 2015 outburst. We have been able to detect the pulsed emission from the source up to the $\sim150$~keV. The standard binary evolution scenario suggests that the orbital period is caused by angular momentum loss through gravitational waves, or by magnetic braking \citep{tauris06}. The AMXP SAX J1808.4--3658 has shown an orbital expansion, but one that is much faster than expected from the standard binary evolution theory \citep[][and references therein]{disalvo08,hartman08,patruno12b}. For \igr\ $\dot{P}_{\rm orb}$ is consistent with zero, at variance of SAX J1808.4--3658 and orbital evolution prediction. The single \swift\ pointings show no evidence of significant spectral variability and the average broad-band (1.1 -- 250 keV) spectrum is best fitted with a thermal Comptonization model. 

We have also reported on the discovery of the first thermonuclear burst emitted by the source and occurred around the peak of the 2015 outburst, roughly 2.5 d after the onset of the event. We have noted, that the type-I burst lasted $\sim 7$ min, which is most similar to the ``intermediate-duration'' type-I bursts, that last between 15 -- 30 min and are powered by pure helium \citep[see, e.g., ][]{intzand05,falanga08,falanga09b}. However, in the \igr\ type-I burst the accretion rate is low enough to build up a thick layer of pure helium, and any accreted hydrogen will be exhausted (via the hot CNO cycle) at the base of the layer within $\sim 10$ hr \citep{lampe16}. In this case we consider that the source accretes hydrogen matter with the solar hydrogen fraction $\langle X\rangle=0.7$ and metallicity similar to solar with mass fraction of CNO elements $Z_{\rm CNO} = 0.02$. The burst recurrence times are comparatively longer to reach ignition because there is no contribution from steady hydrogen burning at the base. This is in agreement with the accumulated mass and the inferred ignition column observed by \igr\ for a helium ignition at low mass accretion rates. It is worth mentioning that the calculated ignition depth is a factor of approximately ten larger than the typical short (tens of seconds bursts) for helium burst in high accretion rate sources, which means that the photon diffusion time is also at least a factor of approximately ten larger. The presence of a possible PRE phase is identified by the double-peaked BAT data of the type-I burst, providing the opportunity to estimate the source distance with a relatively good accuracy at $d=4.2\pm0.5$ kpc. This value is  within the previous estimated source distances ranging from 2 to 4 kpc \citep{galloway05,jonker05,torres08}.

\begin{acknowledgements} 
This research was financed by the Swiss National Science Foundation project 200021\_149865. VdF and MF acknowledge the Department of Physics at the University of Basel and especially Friedrich-K. Thielemann. VDF and MF also thank the International Space Science Institute in Bern for their support.
\end{acknowledgements} 
 
\bibliographystyle{aa}
\bibliography{references} 

\end{document}